\documentclass[english,aps,prb,twocolumn,floats,showpacs]{revtex4}
\usepackage[T1]{fontenc}
\usepackage[latin9]{inputenc}
\usepackage{amsmath}
\usepackage{amssymb}
\usepackage{graphicx}
\usepackage{esint}

\makeatletter

\@ifundefined{textcolor}{}
{%
 \definecolor{BLACK}{gray}{0}
 \definecolor{WHITE}{gray}{1}
 \definecolor{RED}{rgb}{1,0,0}
 \definecolor{GREEN}{rgb}{0,1,0}
 \definecolor{BLUE}{rgb}{0,0,1}
 \definecolor{CYAN}{cmyk}{1,0,0,0}
 \definecolor{MAGENTA}{cmyk}{0,1,0,0}
 \definecolor{YELLOW}{cmyk}{0,0,1,0}
}

\usepackage{babel}
\usepackage{bm}

\makeatother

\usepackage{babel}
\begin{document}

\title{Manipulating Magnetic Moments by Superconducting Currents}

\author{Eugene M. Chudnovsky}

\affiliation{Physics Department, Herbert H. Lehman College and Graduate School, The City University of New York \\
 250 Bedford Park Boulevard West, Bronx, New York 10468-1589, USA}

\date{\today}
\begin{abstract}
We show that the interaction between a superconducting order parameter and the magnetic moment of an atomic cluster in a two-dimensional s-wave superconductor with Rashba spin-orbit coupling generates magnetic anisotropy that can be stronger or comparable to the magnetic anisotropy due to the crystal field and the shape of the cluster. Transport current through the superconductor produces the effective magnetic field acting on the cluster's magnetic moment. The direction of the effective field depends on the direction of the current, thus allowing one to manipulate the magnetic moment by the superconducting current. Due to the large density of the superconducting current this method of magnetization reversal can be more advantageous at low temperatures than the spin-transfer torque method that requires a large spin-polarized current through a normal metal.  
\end{abstract}


\maketitle

There is a growing interest to the interaction of superconducting currents with spins in nanostructures. It has given rise to the field of superconducting spintronics \cite{Eschrig,Linder}. Initial focus had been on proximity effects at the interface of a superconductor and a ferromagnet \cite{Buzdin2005} and on spins inside Josephson junctions \cite{Houzet-PRL08,Petkovic-PRB09,CC-PRB10,Buzdin-PRL08,Buzdin-PRL09,Dolcini-PRB2015,Konschelle-PRB2015,EC-PRB16}. More recently, the search for topological superconductivity and Majorana fermions has ignited theoretical and experimental work on interplay between magnetic moments and superconductivity in one- and two-dimensional (2D) conductors \cite{Choy-PRB11,Leijnse,Braunecker-PRL13,Sato}. Progress in manufacturing 2D superconductors (see, e.g., Refs.\ \onlinecite{Efren,Brun}), specifically the ones with large spin-orbit interaction \cite{Matetskiy,Nam}, led to the discussion of symmetry-breaking persistent currents generated by nanoscale magnetic clusters and magnetic impurities  \cite{Pershoguba-PRL15,Bjornson-PRB15}. 

Driven by computer industry, switching of magnetic moments by means other than applying the magnetic field has been the paradigm of modern magnetism.  Methods based upon application of short electric pulses to multiferroic and composite materials have been explored \cite{Matsukura-NatureNano2015}. Spin-polarized currents have been shown to do the job, which has led to the commercialization of random access memory (STT-RAM) devices \cite{Brataas-NatureMaterials2012}. The necessity to run large currents through thin normal layers limits applications of this method. Superconductors can sustain much greater currents, which, in some cases, could be a trade-off for operating below room temperature. Recently, the possibility of controlling magnetism by superconductivity in FM/SC/FM structures has been demonstrated \cite{Golubov-NatMat2017,Zhu-NatMat2017}. In this paper we are asking whether it is possible to manipulate a nanoscale magnetic moment embedded in a single  superconducting layer by a superconducting current. We demonstrate that 2D superconductors with large spin-orbit coupling present such an opportunity. 

Most of the previous research on interaction between magnetism and superconductivity focused on the effect of a static magnetic order on a superconducting order parameter. This is justified by the fact that typical exchange energies that are responsible for the ferromagnetic order are much greater than pairing interactions responsible for superconductivity. While this is certainly true for magnetic vs superconducting order, the equilibrium orientation of the magnetic moment of a ferromagnet is determined by the magnetic anisotropy that is of relativistic spin-orbit origin. In many cases it is comparable to the superconducting gap. Consequently, it should not come as a surprise that superconductivity can have a profound effect on the orientation of the magnetic moment. 

By computing the additional energy due to persistent currents induced by the magnetic cluster in a 2D superconductor with spin-orbit coupling we show below that the currents induce large easy-plane magnetic anisotropy that forces the magnetic moment of the cluster to lie in the plane of the superconductor. In addition we will demonstrate that a transport current through the superconductor with spin-orbit coupling generates an effective magnetic field in the cluster that is capable of switching the direction of the magnetic moment between two opposite equilibrium orientations along the easy anisotropy axis. This effect is a superconducting member of the family of effects arising from the current-induced spin-orbit torque that have been intensively studied in recent years \cite{Gambardella}.  We also show that combination of a strong easy-plane magnetic anisotropy due to the superconductor with a weak easy-axis anisotropy due to the crystal field or the shape of the cluster may lead to a significant rate of quantum tunneling of the magnetic moment, thus, providing a possible design for a qubit. 

\begin{figure}
\vspace{-2cm}
\includegraphics[width=200mm]{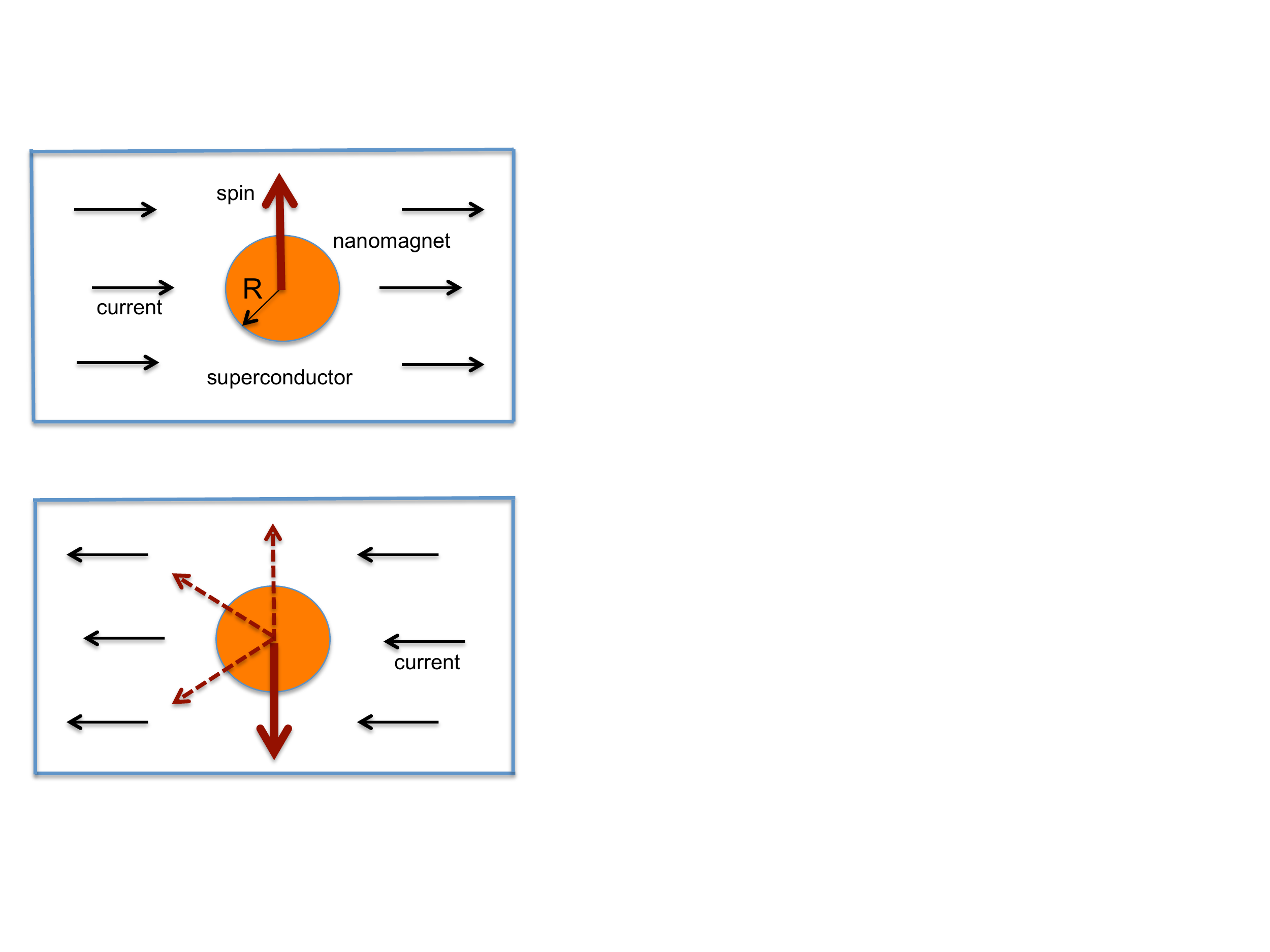}
\vspace{-2.5cm}
\caption{Switching the magnetic moment by reversing the superconducting current.}
\label{geometry}
\end{figure}
As a generic example we consider an s-wave 2D superconducting layer with Rashba spin-orbit coupling 
\begin{equation}
H_{so} =\lambda ({\bm \sigma}\times {\bf p})_z
\end{equation}
in the limit 
\begin{equation}
p_F \gg p_R = m\lambda \gg \frac{\Delta}{\lambda}, 
\end{equation}
with $m$ and $p_F$ being the electron mass and the Fermi momentum, $p_R = m\lambda$ being the Rashba momentum, and $\Delta$ being the superconducting gap. We place a 2D nanoscale ferromagnetic cluster of total spin $S \gg 1$, and of  radius $R$ that is below the superconducting coherence length $\xi$, on the surface of the superconductor as is shown in Fig.\ \ref{geometry}. The condition $S \gg 1$ allows one to treat the spin of the cluster classically. Its exchange interaction with electron spins ${\bm \sigma}$ is assumed to be of a standard form
\begin{equation}
H_{ex} =-\frac{1}{2}JS\int d^2 r \Psi^{\dagger}({\bf r})[{\bf s}({\bf r})\cdot {\bm \sigma}]\Psi({\bf r}),
\label{H-ex}
\end{equation}
with $J$ being the exchange constant and ${\bf s} = {\bf S}/S$ being a classical unit vector in the direction of ${\bf S}$. 

The proximity of a ferromagnet induces a finite spin polarization in the electron fluid. In turn, the spin-orbit interaction, together with a broken inversion symmetry at the surface, generates an additional term in the expression for the superconducting current.  To the first order on $\lambda$ the corresponding superfluid  flow (superconducting current divided by the electron charge) is given by \cite{Pershoguba-PRL15}
\begin{equation}
{\bf j} = \frac{\hbar n_s}{2m}{\bm \nabla} \theta + \alpha(\hat{\bf z}\times{\bf s}),
\label{superflow}
\end{equation}
where  $\theta({\bf r})$ is the phase of the superconducting order parameter, $n_s$ is a 2D density of superconducting electrons, and 
\begin{equation}
\alpha = \left(\frac{ JS}{2\pi \hbar^2}\right)p_R
\end{equation}
is a parameter proportional to both, the strength of the exchange interaction, $J$, and the strength of the spin-orbit interaction, $\lambda p_F$. 

Eq.\ (\ref{superflow}) follows from the minimization of the Ginzburg-Landau free energy amended by the term arising from the spin-orbit coupling,
\begin{equation}
F = \int d^2r\left[\frac{n_s}{2m}\left(\frac{\hbar {\bm \nabla} \theta}{2}\right)^2 + \alpha(\hat{\bf z}\times{\bf s})\cdot \left(\frac{\hbar {\bm \nabla} \theta}{2}\right)\right].
\label{free-energy}
\end{equation}
Writing from Eq.\ (\ref{superflow}) $(\hbar {\bm \nabla}/2) \theta = ({m}/{n_s})\left[{\bf j} - \alpha(\hat{\bf z}\times{\bf s})\right]$ and substituting it into Eq.\ (\ref{free-energy}) one obtains
\begin{equation}
F = \frac{m}{2n_s}\int d^2r \left[{\bf j}^2 - \alpha^2(\hat{\bf z}\times{\bf s})^2 \right].
\label{F}
\end{equation}

The persistent current, ${\bf j} = {\bf j}_s({\bf r})$, generated by the ferromagnetic cluster of radius $R < \xi$ has been computed in Ref.\ \onlinecite{Pershoguba-PRL15} by solving the equation for the phase,
\begin{equation}
\nabla^2\theta = \frac{2m\alpha}{n_s}\hat{\bf z}\cdot\left[{\bm \nabla}\times {\bf s}({\bf r})\right]=-\frac{2m\alpha} 
{n_s}\hat{\bf z}\cdot\left(\hat{\bf r}\times {\bf s}\right)\delta(r-R),
\label{Laplace}
\end{equation}
that follows from the continuity condition ${\bm \nabla}\cdot{\bf j} = 0$. The solution was obtained by using the Green function, $G({\bf r}) = (2\pi)^{-1}\ln(r)$, of the 2D Laplace equation, satisfying $\nabla^2G({\bf r}) =\delta({\bf r})$. Substituting the resulting $\theta({\bf r})$ into Eq.\ (\ref{superflow}), one gets for \cite{Pershoguba-PRL15}
\begin{equation}
{\bf j}_s({\bf r})  =\frac{1}{2}\alpha(\hat{\bf z}\times{\bf s})
\label{j1}
\end{equation}
at $ r < R$ and
\begin{equation}
{\bf j}_s({\bf r})  = \alpha \left(\frac{R}{r}\right)^2\left\{[{\bf r}\cdot(\hat{\bf z}\times{\bf s})]\frac{\bf r}{r^2} - \frac{1}{2}(\hat{\bf z}\times{\bf s})\right\}
\label{j2}
\end{equation}
at $ r > R$. 

Substitution of Eqs. (\ref{j1}) and (\ref{j2}) into Eq. (\ref{F}) gives
\begin{equation}
F = F_s =-\frac{\pi m \alpha^2R^2}{4n_s}(\hat{\bf z}\times{\bf s})^2 = {\rm const} + K_s s_z^2
\label{Ks}
\end{equation}
where we have used $(\hat{\bf z}\times{\bf s})^2 = 1 - s_z^2$. 

Thus, the induction of the persistent currents by the magnetic cluster results in the easy-plane magnetic anisotropy,  $K_s = \pi m \alpha^2R^2/(4n_s)$, that confines the magnetic moment of the cluster to the xy plane of the 2D superconductor.  To estimate the magnitude of this anisotropy it suffices to write $\alpha$ in the form $\alpha = [JS/(2E_F)]\lambda n_e$, where  $E_F = \pi \hbar^2 n_e/m$ is the Fermi energy of a 2D electron gas of density $n_e = \langle \Psi^{\dagger}({\bf r}) \Psi({\bf r})\rangle$.  This gives 
\begin{equation}
K_s = \frac{\pi}{4}m\lambda^2\left(n_e R^2\right)\left(\frac{n_e}{n_s}\right)\left(\frac{JS}{2E_F}\right)^2.
\label{Ks}
\end{equation}
Formally, $K_s$ increases on decreasing $n_s$. However, the critical current decreases on decreasing $n_s$ as well. Since the persistent current cannot exceed the critical current, this means that one can hardly achieve large $K_s$ on approaching the critical temperature. It is, therefore, makes sense to use $n_s \sim n_e$ for the estimate of $K_s$. Choosing $\lambda \sim 10^5$ m/s, $n_e R^2 \sim 10^2$, and $JS \sim 2E_F$, one obtains from Eq.\ (\ref{Ks}) $K_s \sim 10$ eV. This is a very large magnetic anisotropy that should force the magnetic moment of the cluster to lie in the 2D plane. 

Consider now a situation with a transport current ${\bf j}_0$. Since Eq.\ (\ref{Laplace}) allows the addition of $\theta_0({\bf r})$ satisfying ${\bm \nabla}\theta_0 = [2m/(\hbar n_s)]{\bf j}_0 = {\rm const} $ to the solution $\theta({\bf r})$ of  the nonuniform equation, the addition of ${\bf j}_0$ modifies the current as ${\bf j}({\bf r}) = {\bf j}_0 + {\bf j}_s({\bf r})$. Substitution of the total current ${\bf j}({\bf r})$ into Eq.\ (\ref{F}) leads to the addition of the terms to the free energy  that describe the coupling between the spin and the transport current. Working out the integrals one finds that the terms containing logarithmic dependence on the system size cancel out, leaving the following coupling energy
\begin{equation}
F_{js} =\frac{\pi m \alpha R^2}{2n_s}{\bf j}_0\cdot(\hat{\bf z}\times{\bf s}) = -\frac{\pi m \alpha R^2}{2n_s}{\bf s}\cdot(\hat{\bf z}\times{\bf j}_0).
\label{js}
\end{equation}
This equation shows that the effect of the transport current on the magnetic moment in a 2D superconductor with spin-orbit coupling is equivalent to the effect of the magnetic field. Introducing the magnetic moment of the cluster, ${\bf M} = g_e\mu_B {\bf S}$, we have $F_{js} = - {\bf M}\cdot {\bf B}_{j}$ with
\begin{equation}
{\bf B}_{j} = \frac{\pi p_R }{4g_e\mu_B}(n_e R^2)\left(\frac{J}{E_F}\right)(\hat{\bf z}\times{\bf v}_0),
\label{B}
\end{equation} 
where $\mu_B$ is the Bohr magneton, $g_e$ is the gyromagnetic factor, and ${\bf v}_0 ={\bf j}_0/n_s$ is the drift velocity of the electron fluid in the superconducting current. 

Normally the ferromagnet would have an easy magnetization axis in the xy plane, say the y-axis for certainty. Let us begin with the magnetization of the cluster looking in the positive y-direction. Switching of the magnetic moment by the superconducting current is illustrated in Fig.\ \ref{geometry}. If the current ${\bf j}_0$ runs in the positive x-direction the magnetic field ${\bf B}_{j}$ it generates due to the spin-orbit coupling looks in the y-direction, that is, it is aligned with the magnetization of the cluster. If the current is now reversed, ${\bf j}_0 \rightarrow -{\bf j}_0$, and begins to flow in the negative x-direction, so does ${\bf B}_{j}$ that is now looking in the negative y-direction. This must lead to the magnetization reversal if ${B}_{j}$ is sufficiently large to overcome the energy barrier due to magnetic anisotropy. 

The magnetic energy can be written as \cite{Lectures}
\begin{equation}
F_{M} = \int d^2 r \left[-\frac{1}{2}\beta_{\parallel}M_{0y}^2 + \frac{1}{2}\beta_{\perp}M_{0z}^2 -{\bf M}_0\cdot{\bf B}_{j}
\label{M}\right]
\end{equation}
in terms of the 2D magnetization ${\bf M}_0=g_e\mu_B {\bf S}/(\pi R^2)$. The first term in Eq.\ (\ref{M}), with $\beta_{\parallel} > 0$, represents the easy axis magnetic anisotropy in the xy plane, while the second term with $\beta_{\perp} = 2K_s/(g_e\mu_B S)^2$ represents the easy plane anisotropy generated by persistent currents flowing around the magnet. The effective magnetic field acting on the magnetic moment is
\begin{equation}
{\bf B}_{\rm eff} = -\frac{\delta F_M}{\delta {\bf M}_0} = \beta_{\parallel} M_0\hat{\bf y} - \beta_{\perp}M_0\hat{\bf z} + {\bf B}_{j}.
\label{eff}
\end{equation}
Strong easy-plane anisotropy makes equilibrium ${\bf M}_0$ to lie in the xy plane. The current ${\bf j}_0 =\pm  j_0\hat{\bf x}$ generates ${\bf B}_{j}= \pm {B}_{j}\hat{\bf y}$. According to Eq.\ (\ref{eff}) the reversal of ${\bf M}_0$ from the positive  $y$-direction to the negative $y$-direction requires ${\bf B}_{j} = -B_{j}\hat{\bf y}$ with $B_{j} > B_{\parallel} = \beta_{\parallel}M_0$. The field $\beta_{\parallel}M_0$ is called the anisotropy field. For the majority of ferromagnets it  lies between $0.01$ T and $0.1$ T.

To estimate whether a superconducting current can generate the effective spin-orbit magnetic field in the magnet that is comparable to the anisotropy field we first notice that according to Eq.\ (\ref{B}) ${B}_j$ depends neither on $S$ nor on $n_s$. Independence from $n_s$ may create the wrong impression that our results equally apply to the case of $n_s = 0$, that is, to a normal conductor. However, our formulas rely on the assumptions that the currents are non-dissipative, given by Eq.\ (\ref{superflow}), and that $R < \xi$, which justifies the validity of Eq.\ (\ref{j1}) and can only be true for a superconductor. When materials of the ferromagnetic cluster and of the superconductor are chosen the only free parameters in the expression for $B_j$ are the size of the magnetic cluster $R$ and the electron drift velocity ${\bf v}_0$ that can be very large in a superconductor as compared to normal metals. At $n_e R^2 \sim 10^2$ and $J \sim 10^{-2}E_F$, for a sufficiently large spin-orbit coupling, $\lambda \sim 10^5$ m/s, and large current density, $v_0 \sim 10$ m/s (typically corresponding to $0.1$ A/cm for $n_s \sim 10^{15}$ cm$^{-2}$), one obtains from Eq.\ (\ref{B}) $B_{j} \sim 0.1$ T. This demonstrates a principal possibility to reverse the magnetic moment of a nanoscale ferromagnetic cluster by the current through a 2D superconducting layer with strong spin-orbit coupling. The advantage of a superconductor is that the required current density is easier to achieve  than the current needed to induce magnetization reversal by a spin-transfer torque in a normal metal. 

Finally, we would like to comment on the possibility of using spins in a 2D superconductor with spin-orbit coupling as qubits. To have a functional qubit one needs a quantum superposition of spin-up and spin-down states that survives the decohering effect of the environment long enough to permit computation. For practical purposes it is also desirable to have a qubit with a sufficiently large magnetic moment that one can easily measure. This means a large total spin, $S \gg 1$, that can tunnel fast between up and down orientations. Generally speaking the conditions of large $S$ and large decoherence time are at odds with each other because the WKB action associated with the spin \cite{MQT-book} is proportional  to the total spin $S$, thus making the spin tunneling rate to decrease exponentially with increasing $S$. However, in the problem discussed above the possibility of a large tunneling rate for a large spin is provided by the large easy plane anisotropy generated by non-dissipative currents of spin-orbit origin flowing around the spin, see Eq.\ (\ref{Ks}) and discussion below it. At $\beta_{\perp} \gg \beta_{\parallel}$ the enhancement of tunneling between opposite orientations of the magnetic moment along the easy y-axis comes from the large second term in Eq.\ (\ref{M}) that does not commute with the first term when components of ${\bf M}_0$ are treated as quantum operators. The tunneling rate is proportional to $\exp(-U/T_Q)$, where $U$ is the energy barrier due to the magnetic anisotropy and $T_Q \propto [\beta_{\parallel}(\beta_{\parallel}+\beta_{\perp}]$ is the characteristic temperature of quantum fluctuations \cite{MQT-book}. At $\beta_{\parallel} \sim \beta_{\perp}$ the WKB tunneling exponent is of the order of the tunneling spin $S$, which makes the tunneling rate  exponentially small for large $S$. However, in the limit of $\beta_{\parallel} \ll \beta_{\perp}$ the tunneling exponent is $2S(\beta_{\parallel}/\beta_{\perp})^{1/2}$, thus allowing a significant rate of tunneling for $S \gg 1$. The tunneling rate can be further increased by reducing the barrier $U$ with the help of the transport current. 

In conclusion, we have shown that currents induced by a magnetic cluster in a 2D superconducting layer with spin-orbit coupling generate significant magnetic anisotropy that affects equilibrium orientation of the magnetic moment of the cluster. In addition, the transport current in a superconductor generates the effective magnetic field on the cluster that can switch the orientation of its magnetic moment. This provides an interesting mechanism of magnetization switching by a current, which is conceptually different from the switching by a spin-transfer torque. While the latter relies on a large exchange interaction of the spin-polarized normal current with ferromagnetic spins, the mechanism we propose relies on a large spin-orbit interaction and large densities of the transport current that can be achieved in superconductors. We also show that the proposed mechanism can be used to achieve a significant rate of quantum tunneling of the magnetic moment in a magnetic cluster deposited on the surface of a superconductor. 

This work has been supported by the grant No. DE-FG02-93ER45487 funded by the U.S. Department of Energy, Office of Science.

\end{document}